\documentclass[journal]{IEEEtran}

\usepackage[cmex10]{amsmath}
\usepackage{amssymb}
\usepackage{amsfonts}
\usepackage{amsthm}

\usepackage{color}
\usepackage{cite}

\usepackage{enumerate}

\usepackage{graphicx}

\usepackage{multirow}

\usepackage{subfigure}

\usepackage{url}

\def \N        {\hbox{I \kern-.5em N}}
\def \R        {\hbox{I \kern-.5em R}} 
\def \Q        {\hbox{\hspace*{.25em}{\sf I} \kern-.83em Q}}
\def \ZZ       {\hbox{\sf  Z \kern-.8em Z}}

\def \dual      {u}
\def \ModelName {DFOP} 

\def \EdgeIdx   {e}

\def \ILP       {\textsc{ilp}}

\def \LinkSet   {L}
\def \LinkIdx   {\ell}
\def \LP        {\textsc{lp}}

\def \NodeSet   {V}
\def \NodeIdx   {v}

\def \subnetIdx {\textsc{fsn}}
\def \subnetSet {\mathcal{FSN}}

\def \varkFSN   {b_k^{\subnetIdx}}

\def \xColor    {x_{\lambda}}

\def \zFSN      {z_{\subnetIdx}}

\begin{document}
%
\title{A Two Sub-problem Decomposition for the \\ Optimal Design of Filterless Optical Networks}
%
%
%

\author{Brigitte Jaumard,~\IEEEmembership{Senior Member,~IEEE,}
    and Yan Wang
\thanks{B. Jaumard and Y. Wang were with the Department
of Computer Science and Software Engineering, Concordia University, Montreal,
QC, H3G 1M8 Canada e-mail: brigitte.jaumard@concordia.ca}
\thanks{Manuscript received December 31st, 2020}}

%
%

\markboth{Jaumard \MakeLowercase{\textit{et al.}} A Two Sub-problem Decomposition for the Optimal Design of Filterless Optical Networks}%
{Jaumard \MakeLowercase{\textit{et al.}}: Bare Demo of IEEEtran.cls for IEEE Journals}
%



\maketitle

\begin{abstract}
Filterless optical transport networks relies on passive optical interconnections between nodes, i.e., on splitters/couplers and amplifiers. 
While different studies have investigated their design, none of them offer a solution for an optimal design. 
We propose a one step solution scheme, which combines network provisioning, i.e., routing  and  wavelength  assignment  within a single mathematical  model.
Decomposition into two different types sub-problems is then used in order to conceive an exact solution scheme. The first type of subproblem relies on the generation of filterless subnetworks while the second one takes care of their wavelength assignment.

Numerical experiments demonstrate the improved performance of the proposed optimization model and algorithm over the state of the art, with the improvement of the solution for several open source data sets.
\end{abstract}


%
\IEEEpeerreviewmaketitle

\section{Introduction}

The idea of filterless optical networks goes back to the seminal articles of \cite{tza99,sav10}. 
These networks rely on broadcast-and-select nodes equipped with coherent transceivers, as opposed to active switching networks, which today use Reconfigurable Optical Add-Drop Multiplexers (ROADMs).
Filterless optical networks exhibit several advantages and are currently considered for, e.g., metro regional networks \cite{pav20} or submarine networks \cite{noo17}.
They allow a reduction in energy consumption and therefore lead to both a reduction in cost and in the carbon footprint.
In addition, they improve the network robustness, while simplifying several aspects of impairment-aware design \cite{tre17}, although with a reduced spectral efficiency due to their inherent channel broadcasting which may be attenuated with the use of blockers, see, e.g., Dochhan \textit{et al.} \cite{doc19}.

In order to solve the filterless network design problem, Tremblay \textit{et al.} \cite{tre13} proposed a two-step solution process: 
(1) A Genetic Algorithm to generate fiber trees, 
(2) A Tabu Search algorithm for solving  the routing and wavelength assignment algorithm on the selected trees among those generated by step (1). Both steps are solved with a heuristic, while efficient exact algorithms exist today, at least for the routing and wavelength assignment problem, see, e.g., \cite{duh16,jau17_TON}.

Ayoub \textit{et al.} \cite{ayo18} also devised a two-step algorithm:
(1) A first heuristic algorithm to generate edge-disjoint fiber trees,
(2) An ILP (Integer Linear Program) model to solve the routing and spectrum assignment problem. Unfortunately, they did not compare their results with those of \cite{tre13}.

In this paper, we improve on the solution process we had previously proposed in \cite {jau18}, in which we had an exact algorithm to generate fiber trees, but we were still with a two-step process in which the second step, as in previous studies, solves the routing and wavelength assignment problem. 
We are now able to propose a first exact one-step solution process using a large scale optimization modelling and algorithm, which are described in Sections \ref{sec:math_model} and \ref{sec:sol_scheme}, respectively. Numerical results then show improved filterless network designs over all the previous studies.

The paper is organized as follows. The problem statement of the design of filterless optical networks is recalled in Section \ref{sec:design}. 
In Section \ref{sec:math_model}, the mathematical model of our one-step design is proposed. In Section \ref{sec:sol_scheme}, we discuss our solution scheme: generation of an initial solution, column generation technique to solve the linear relaxation and derivation of an integer solution. Computational results are summarized in Section \ref{sec:results}, including a comparison with the results of \cite{tre13}. Conclusions are drawn in the last section.


\section{Design of Filterless Optical Networks}
\label{sec:design}

Consider an optical network represented by its physical network $G = (V,L)$, where $V$ is the set of nodes (indexed by $v$), and $L$ is the set of fiber links (indexed by $\ell$). We denote by $\textsc{in}(v)$ and $\textsc{out}(v)$ the set of incoming and outgoing links of $v$, respectively.

In the sequel, we will use undirected trees and will adopt the following terminology: we use the term "edge" for designating an undirected link, and the term link for designating a directed link. We denote by $\omega(v)$ the co-cycle of $v$, i.e., the set of adjacent edges of $v$ in an undirected graph.

The traffic is described by a set $K$ of unit requests where each request $k$ is described by its source and destination nodes, $v_s$ and $v_d$ respectively.

\begin{figure}[htb]
  \begin{center}
  \subfigure[Undirected graph supporting the physical network]
  {\includegraphics[width =.48\columnwidth]{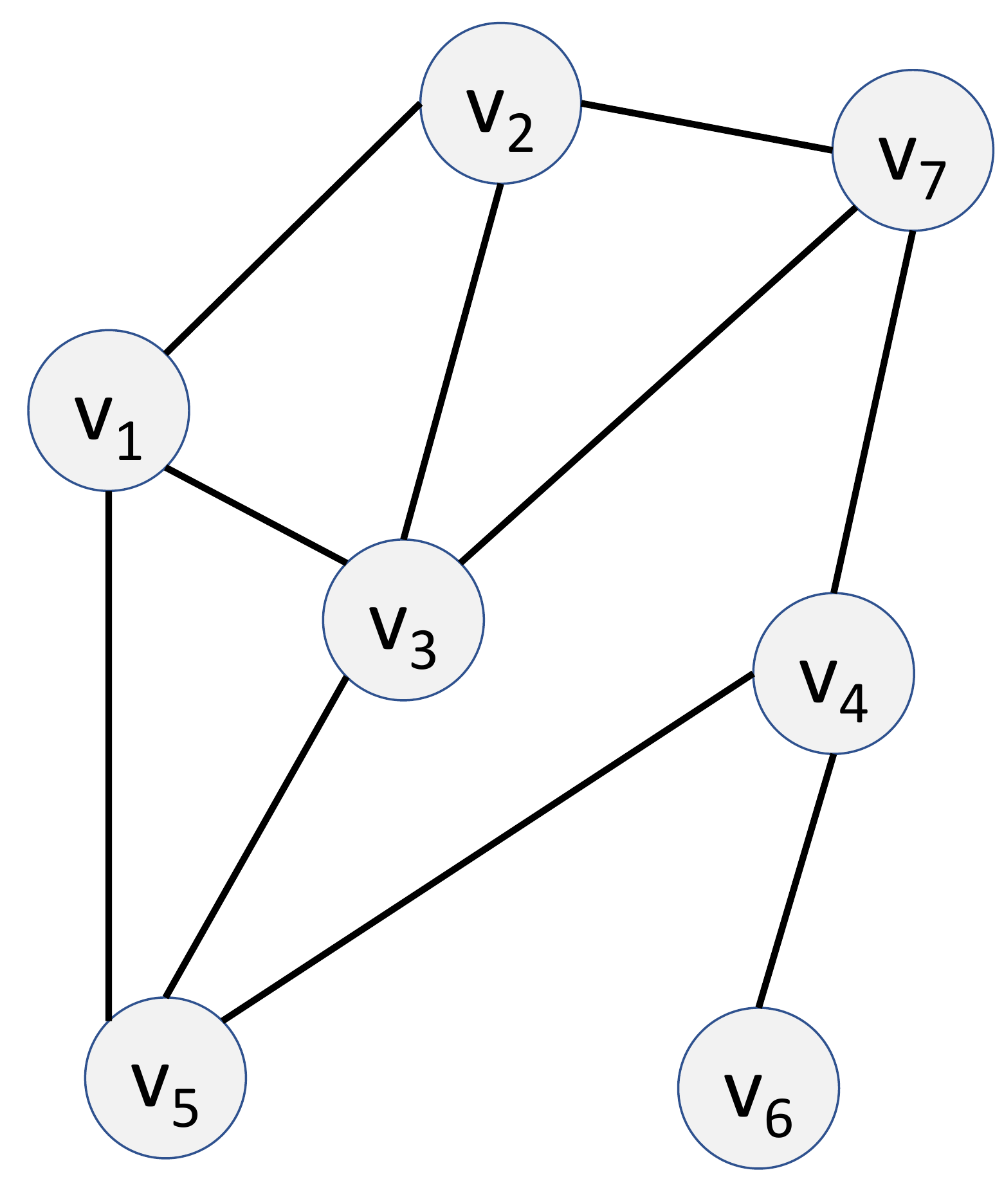}\label{fig:fig1a}}
  \subfigure[Three different FSNs]
  {\includegraphics[width = .48\columnwidth]{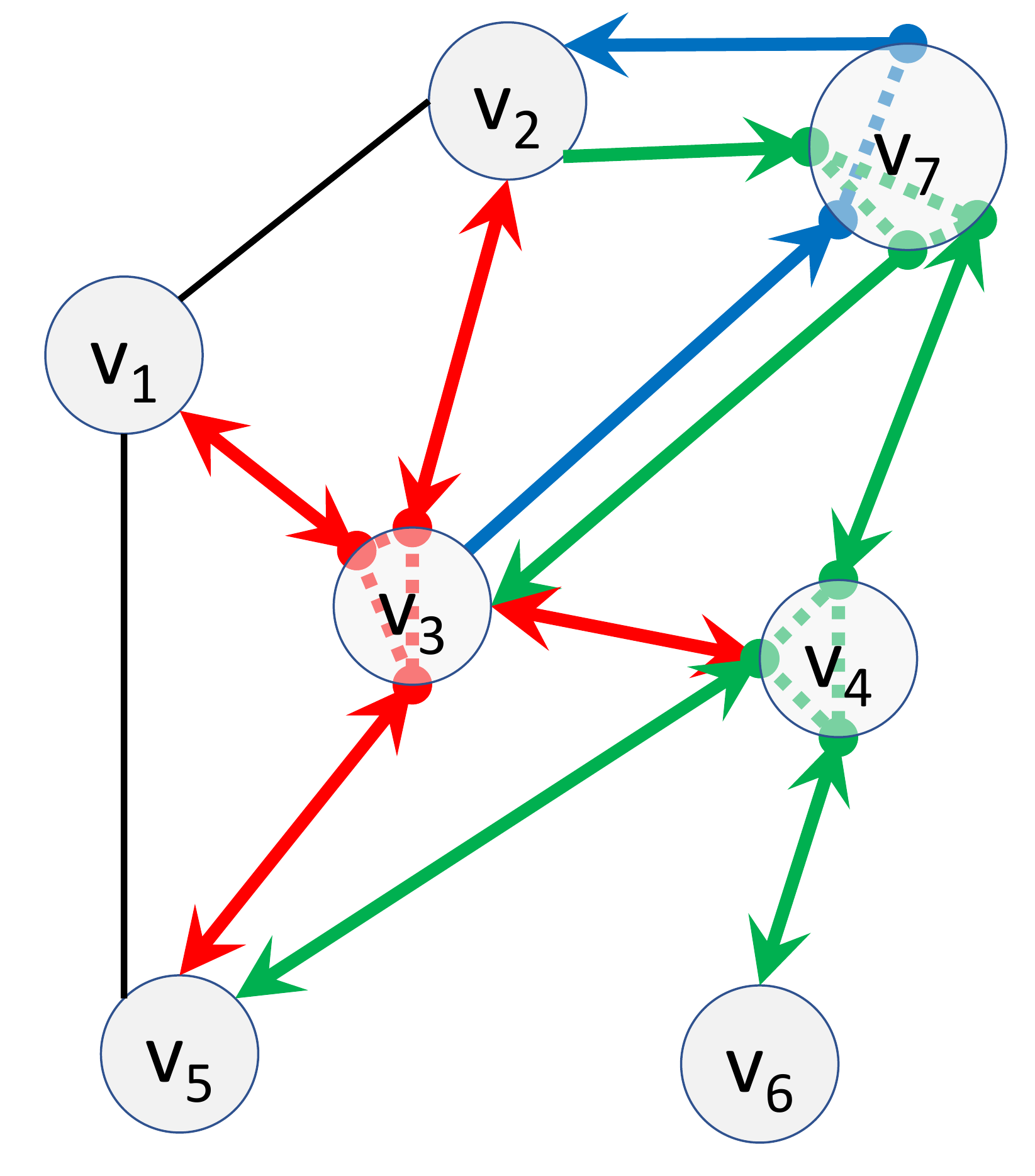}\label{fig:fig1b}}
  \end{center}
  \caption{Construction of a filterless network solution}
  \label{fig:example1}
\end{figure}

A filterless network solution consists of a set of filterless sub-network (f-subnet or FSN for short) solutions, where each f-subnet is a directed graph supported by an undirected tree, and f-subnets are pairwise fiber links disjoint.
We provide an illustration in Figure \ref{fig:example1}. 
Figure \ref{fig:fig1a} represents the node connectivity of 
the original optical network.   We assume that each pair of connecting nodes has two directed fibers, one in each direction.
Figure \ref{fig:fig1b} depicts filterless optical network with three f-subnets, each supported by an undirected tree:
the red one   (FSN$_1$), supported by tree $T_1 = \{ \{v_1, v_3\}, \{v_2, v_3\},\{v_3, v_4\},\{v_3, v_5\}\}$,
the blue one  (FSN$_2$), supported by tree $T_2 = \{ \{v_3, v_7\}, \{v_2, v_7\}\}$ and
the green one (FSN$_3$), supported by tree $T_3 = \{ \{v_3, v_7\}, \{v_2, v_7\},\{v_4, v_7\},\{v_4, v_5\},\{v_4, v_6\}\}$. 
Fiber links with two arrows indicates the filterless optical network uses two directional fibers, one in each direction, while links with a unique arrow use each a single directional fiber. 
Required passive splitters and combiners are added to interconnect the fiber links supporting the request provisioning, see, e.g., node $v_4$ requires a three splitters/combiners with two ports in order to accommodate the provisioning of the requests using the green filterless network.
Note that all three sub-networks are pairwise fiber link disjoint.
Observe that two node connections are unused: edges $\{ v_1, v_2 \}$ and $\{ v_1, v_5 \}$ of the physical network, which may be an indication of a poor network design. 

Each request is provisioned on one of the sub-networks. 
 Observe that if we use more than one f-subnet, the routing is not necessarily unique. For instance, in the example of Figure \ref{fig:fig1b}, request from $v_2$ to $v_3$ can be either provisioned on the FSN$_1$ (red) filterless sub-network, or on FSN$_3$ (green sub-network).

We are interested in establishing a set of \underline{fiber link} disjoint f-subnets such that
the provisioning of all the requests minimizes the maximum fiber link load. 
The objective is to find f-subnets and a provisioning such that we minimize the overall number of wavelengths. 
Each request must be provisioning on a lightpath, i.e., a route and a wavelength such that the continuity constraint is satisfied, i.e., the same wavelength from source to destination, while considering the broadcast effect, i.e., same wavelength on all outgoing links of the source node and their descendants.

\begin{figure}[htb]
  \begin{center}    
        \includegraphics[width = .7\columnwidth]{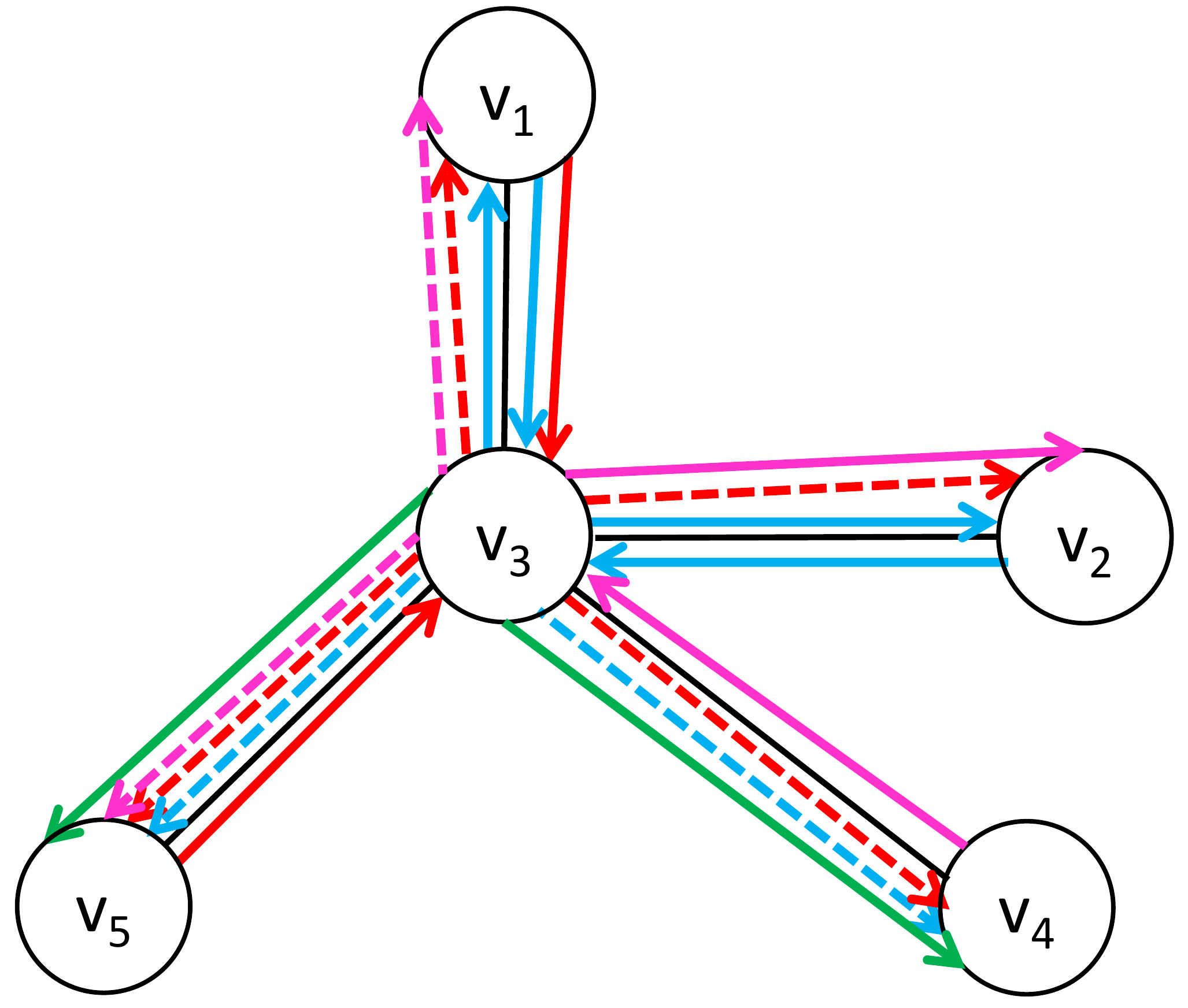}
  \end{center}
  \caption{Wavelength request provisioning}
  \label{fig:WA}
\end{figure}

We next illustrate the impacts of the broadcast effect on the wavelength assignment. 
Consider the following example  with 7 requests on FSN$_1$ in Figure \ref{fig:fig1b}:
$k_{13}$ from $v_1$ to $v_3$, and with the same notations, $k_{53}$, 
$k_{35}$, $k_{21}$, $k_{12}$, $k_{42}$, and $k_{34}$.
Figure \ref{fig:WA} depicts a provisioning of them, which requires 4 wavelengths. Plain lines correspond to the provisioning between the source and the destination nodes, while dotted lines indicate the broadcast effect beyond the destination nodes.
Observe that the wavelength assignment takes into account that requests that conflict only in their broadcast effect beyond their destination nodes can share the same wavelength. 
For instance, request $k_{13}$ can share the same wavelength as request $k_{53}$. 


\section{Mathematical Model}
\label{sec:math_model}


We propose a mathematical model, called {\ModelName}, which relies on two different sets of configurations: the FSN configurations and the wavelength configurations.
Each FSN configuration consists of:
\textit{(i)} a subgraph $\subnetIdx$ of $G$ supported by an undirected tree backbone, and
\textit{(ii)} a set of requests provisioned on $\subnetIdx$. 
Each wavelength configuration consists of a potential wavelength assignment for the overall set of requests.

The design of a filterless network amounts then to the selection of a set of FSN configurations, as many as the desired number of FSNs, and of wavelength configurations, as many as the number of required wavelengths, which jointly minimizes the number of required wavelengths.

The proposed decomposition consists of a master problem coordinating a set of pricing problems which are solved alternately, see Section \ref{sec:sol_scheme} for the details of the solution scheme.
We next describe the master problem and the pricing problems.


\subsection{Master Problem}
\label{sec:Master_Pb}

Let $\subnetIdx$ be a f-subnet configuration. It is formally characterized by:
\begin{itemize}
    \item $a_{\ell}^{\subnetIdx} = 1$ if link $\ell$ is in f-subnet $\subnetIdx$, 0 otherwise
    \item $\varkFSN = 1$ if request $k$ is routed on $\subnetIdx$, 0 otherwise
    \item $\theta^{\subnetIdx}_{kk'} = 1$ if request $k$ and request $k'$ are in conflict on $\subnetIdx$, 0 otherwise.
\end{itemize}

Let $\lambda$ be a wavelength configuration. It is characterized by:
\begin{itemize}
    \item $\beta^{\lambda}_k = 1$ if request $k$ uses wavelength (color) configuration $\lambda$, 0 otherwise.
\end{itemize}    

\noindent
There are two sets of decision variables, one for each set of configurations:
\begin{itemize}
    \item $\zFSN = 1$ if f-subnet FSN, together with its provisioned requests, is selected as a filterless sub-network, 0 otherwise.
    \item $\xColor = 1$ if wavelength (color) configuration $\lambda$ is used, 0 otherwise. One wavelength configuration is defined by all the requests with the same wavelength assignment.
\end{itemize}    


The objective corresponds to the minimization of the total number of wavelengths.
\begin{equation}
    \min \sum\limits_{\lambda \in \Lambda} \xColor \label{eq:obj_Master}
\end{equation}
subject to:
\begin{alignat}{2}
& \sum_{\subnetIdx \in \subnetSet} \varkFSN \zFSN \geq 1
&& k \in K\label{eq:demand} \\
& \sum\limits_{\subnetIdx \in \subnetSet} a_{\ell}^{\subnetIdx} \zFSN \leq 1 \qquad 
&&  \ell \in L \label{eq:link_disjoint} \\
& \sum\limits_{\subnetIdx \in \subnetSet} \zFSN \leq n_\subnetSet 
&& \label{eq:limit_nb_trees} \\
& \sum\limits_{\lambda \in \Lambda} \beta^{\lambda}_k \> \beta^{\lambda}_{k'} \xColor \leq 1 -
  \sum\limits_{\subnetIdx \in \subnetSet} \theta^{\subnetIdx}_{k k'} \zFSN \qquad
&& k, k' \in K \label{eq:conflict_color} \\
& \sum\limits_{\lambda \in \Lambda} \beta^{\lambda}_k x_{\lambda} \geq 1
&& k \in K \label{eq:one_request_at_least_one_color} \\
& \zFSN \in \{0,1\} && \hspace*{-1.cm} \subnetIdx \in \subnetSet \label{eq:variable_z}\\
& \xColor \in \{0,1\} && \hspace*{-1.cm} \lambda \in \Lambda. \label{eq:variable_x}
\end{alignat}
Constraints \eqref{eq:demand} indicate each request will be routed on at least one selected filterless sub-network. 
Although the constraints could be written as equality constraints, inequalities make the model easier to solve in practice for LP (Linear Program)/ILP (Integer Linear Program) solvers. Due to the objective, a request will not be routed more than once, unless there is enough spare capacity to do so, and then we can easily eliminate superfluous provisioning in a post-processing step.
Constraints \eqref{eq:link_disjoint} guarantee pairwise link disjoint sub-networks. 
Constraints \eqref{eq:limit_nb_trees} is an optional constraint to express that we want to limit the number of f-subnet configurations. Constraints \eqref{eq:conflict_color} guarantee the conflicted requests could not share the same wavelength (color). 
Constraints \eqref{eq:one_request_at_least_one_color} express that each request will use at least one wavelength (color). Again inequality is justified as for constraints 
\eqref{eq:demand}, and a post-processing can easily eliminate a double wavelength assignment. Post-processing is justified in view of the enhanced convergence of the solution scheme thanks to the inequalities.


\subsection{PP$_{\subnetIdx}$: FSN Pricing Problem}
\label{sec:FSN_Pricing}

FSN pricing problem aims to find a new FSN configuration which could potentially improve the objective of the continuous relaxation of the restricted master problem, 

\noindent
Variables of the pricing problem directly or indirectly defines the coefficients of the column of the master problem, and are defined as follows:
\begin{itemize}
    \item $a_{\LinkIdx} = 1$ if link $\LinkIdx$ is in the f-subnet under construction, 0 otherwise, for all $\LinkIdx \in \LinkSet$ 
    \item $a_{\NodeIdx} = 1$ if node $\NodeIdx$ belongs to f-subnet under construction, 0 otherwise, for all $\NodeIdx \in \NodeSet$. Remember that f-subnets are usually not supported by a spanning tree.  
    \item $\alpha_{e} = 1$ if edge $e$ belongs to the tree supporting the f-subnet under construction, 0 otherwise, for all $e \in E$.
    \item $x_{k} = 1$ if request $k$ is routed on the f-subnet under construction, 0 otherwise, for all $k \in K, t \in T$
    \item $\varphi_{k \LinkIdx} = 1$ if the routing of request $k$ goes through link $\LinkIdx$, or if the channel used in the routing of request $k$ propagates on link $\LinkIdx$ because it is not filtered, 0 otherwise, for all $\LinkIdx \in \LinkSet, k \in K$.
    \item $\psi_{k \LinkIdx} = 1$ if the routing of request $k$ goes through link $\LinkIdx$ between its source and destination, 0 otherwise, for all $\LinkIdx \in \LinkSet, k \in K$. We need both variables $\varphi_{k \LinkIdx}$ and $\psi_{k \LinkIdx} = 1$ in order to identify the unfiltered wavelength links.
    \item $\theta_{kk'} = 1$ if request $k$ and request $k'$ are in conflict on the f-subnet under construction, 0 otherwise.
    \item $\omega_{k k'\ell}  = \psi^{\subnetIdx}_{k \ell} \varphi^{\subnetIdx}_{k' \ell}$, for all $\LinkIdx \in \LinkSet, k, k' \in K$. In other words, with $\omega_{k k'\ell} = 1$ identifying a link on which $k$ and $k'$ are conflicting either between their source and destination nodes, or with one of them being routed to the broadcast of the other request, and 0 otherwise (no conflict).
\end{itemize}

\noindent
\textbf{Objective}: it corresponds to the reduced cost of variable $\zFSN$, i.e., the objective function of the pricing problems, i.e., the decomposition sub-problems. The reader who is not familiar with linear programming concepts, and in particular with the concept of reduced cost is referred to the book of Chvatal \cite{chv83}. The coefficient 0 indicates that the variable $\zFSN$ does not appear in the objective function \ref{eq:obj_Master} of the master problem, and therefore its coefficient contribution to the reduced cost is null.

\begin{multline}
\min \>  0 
- \sum\limits_{k \in K} \dual_k^{(\ref{eq:demand})} x^k 
+ \sum\limits_{\ell \in L} \dual_{\ell}^{\eqref{eq:link_disjoint}} {\alpha_{\ell}}
+ \dual^{\eqref{eq:limit_nb_trees}}
\\
+ \sum\limits_{k, k' \in K} \dual_{k k'}^{\eqref{eq:conflict_color}} \theta_{kk'}
\end{multline}

We now describe the set of constraints.
\begin{alignat}{2}
\intertext{\color{blue} Building an undirected tree}
& \sum\limits_{\substack{e = \{v, v'\} \in E: \\ v, v' \in \NodeSet'}} {\alpha_e} \leq \vert \NodeSet' \vert -1 \quad 
&& \hspace*{-1.cm} \NodeSet' \subset \NodeSet, \vert \NodeSet'\vert  \geq 3 \label{eq:subtour} \\
& \sum\limits_{\NodeIdx \in \NodeSet}a_{\NodeIdx} =  \sum\limits_{e \in E} {\alpha_e} + 1 
&& \label{eq:single_tree} \\
& 2 \alpha_e \leq a_{\NodeIdx} + a_{\NodeIdx'} 
&& \hspace*{-1.cm} v, v' \in \NodeSet, {e = \{v, v'\}}  \label{eq:edge_exists_if}\\
& \sum\limits_{e \in \omega(\NodeIdx)} \alpha_e \geq a_{\NodeIdx}
&& \hspace*{-1.cm}  \NodeIdx \in \NodeSet \label{eq:node_belongs_if} \\
& a_{\ell} \leq \alpha_e, \> a_{\overline{\ell}} \leq \alpha_e
&& \hspace*{-1.cm} \ell = (\NodeIdx, \NodeIdx'), \overline{\LinkIdx} = (\NodeIdx', \NodeIdx) \nonumber \\
& && \hspace*{-1.cm}   v, v' \in \NodeSet, {e = \{v, v'\}} \label{eq:links_vs_edge} \\
\intertext{\color{blue} Routing of the provisioned requests}
& {\varphi}_{k\LinkIdx} \leq a_{\LinkIdx}  
&&  k \in K,  \LinkIdx \in \LinkSet \label{eq:request_link1} \\
& a_{\LinkIdx} \leq \sum\limits_{k \in K}  {\varphi}_{k\LinkIdx} 
&&  \LinkIdx \in \LinkSet \label{eq:request_link2} \\
& {\varphi}_{k \ell} + {\varphi}_{k \overline{\LinkIdx}} \leq 1 \qquad 
&& \ell = (\NodeIdx, \NodeIdx'), \nonumber \\
& && \hspace*{-1.5cm} \overline{\LinkIdx} = (\NodeIdx', \NodeIdx): v \in \NodeSet, v' \in \NodeSet \label{eq:loop} \\
&    \sum\limits_{\LinkIdx \in {\textsc{in}(d_k)}} {\varphi}_{k\LinkIdx} 
  =  \sum\limits_{\LinkIdx \in {\textsc{out}(s_k)}}{\varphi}_{k\LinkIdx} 
  =  x_{k}
\qquad
&& k \in K \label{eq:flow1} \\
& \sum\limits_{\LinkIdx \in {\textsc{in}(s_k)}}{\varphi}_{k\LinkIdx} =  0 
&& k \in K \label{eq:flow2} \\
\intertext{\color{blue} Propagation of unfiltered channels}
& {\varphi}_{k\LinkIdx'} \leq \sum\limits_{\LinkIdx \in {\textsc{in}(v)}} {\varphi}_{k\LinkIdx}  + 1 - a_{\LinkIdx'}
\qquad
&& k \in K,  \nonumber \\
& && \hspace*{-2.5cm} v \in \NodeSet \setminus \{ {s_k} \}, \LinkIdx' \in {\textsc{out}(v)} \label{eq:unfiltered1} \\
& {\varphi}_{k\LinkIdx} \leq {\varphi}_{k\LinkIdx'} + 2 - a_{\LinkIdx} -  a_{\LinkIdx'}
&& k \in K, \nonumber \\
& && \hspace*{-2.5cm}  v \in \NodeSet, \LinkIdx \in {\textsc{in}(v)}, \LinkIdx' \in {\textsc{out}(v)} \label{eq:unfiltered2} \\
\intertext{\color{blue} "Filtered" provisioning of requests}
& \sum\limits_{\LinkIdx \in {\textsc{in}(d_k)}} {\psi}_{k\LinkIdx} =  \sum\limits_{\LinkIdx \in {\textsc{out}(s_k)}}{\psi}_{k\LinkIdx} =  x_k  \qquad
&& k \in K \label{eq:flowPsi1}\\
& \sum\limits_{\LinkIdx \in {\textsc{in}(\NodeIdx)}}  {\psi}_{k\LinkIdx} =  
  \sum\limits_{\LinkIdx \in {\textsc{out}(\NodeIdx)}} {\psi}_{k\LinkIdx} \leq x_k 
&& k \in K, \nonumber \\
& &&  \NodeIdx \in \NodeSet \setminus ({s_k, d_k}) \label{eq:flowPsi2}  \\
& \sum\limits_{\LinkIdx \in {\textsc{out}(d_k)}} {\psi}_{k\LinkIdx} =  \sum\limits_{\LinkIdx \in {\textsc{in}(s_k)}}{\psi}_{k\LinkIdx} =  0  \qquad
&& k \in K \label{eq:flowPsi3}\\
\intertext{\color{blue} Reach distance}
& \sum\limits_{\ell \in L} \textsc{dist}_{\ell} {\psi}_{k\LinkIdx} \leq \textsc{reach\_dist}
&& k \in K \label{eq:variables_psy}\\
\intertext{\color{blue} Identifying wavelength conflicting lightpaths}
& {\psi}_{k\LinkIdx} \leq \varphi_{k\LinkIdx}  
&&  k \in K,  \LinkIdx \in \LinkSet \label{eq:phi_psi} \\
& \theta_{kk'} \geq \psi^{\subnetIdx}_{k \ell} + \psi^{\subnetIdx}_{k' \ell} - 1 \qquad
&&  \ell \in L, \lambda \in \Lambda, \nonumber \\
& &&  k, k' \in K \label{eq:conflict1} \\
& \theta_{kk'} \geq \psi^{\subnetIdx}_{k \ell} + \varphi^{\subnetIdx}_{k' \ell} - 1
&&  \ell \in L, \lambda \in \Lambda, \nonumber \\
& && k, k' \in K \label{eq:conflict2} \\
& \theta_{kk'} \geq \psi^{\subnetIdx}_{k' \ell} + \varphi^{\subnetIdx}_{k \ell} - 1
&&  \ell \in L, \lambda \in \Lambda, \nonumber \\
& &&  k, k' \in K \label{eq:conflict3} \\
& \theta_{kk'} \leq \sum\limits_{\LinkIdx \in \LinkSet} (\omega_{k k' \ell} + \omega_{k' k \ell})
&&  k, k' \in K \label{eq_full_alt:force_conflict1} \\
& \omega_{k k' \ell} \leq \psi^{\subnetIdx}_{k \ell}
&& \ell \in L,  k, k' \in K \label{eq_full_alt:force_conflict2} \\
& \omega_{k k' \ell} \leq \varphi^{\subnetIdx}_{k' \ell}
&& \ell \in L,  k, k' \in K \label{eq_full_alt:force_conflict3} \\
& \omega_{k k' \ell} \geq \psi^{\subnetIdx}_{k \ell}+\varphi^{\subnetIdx}_{k' \ell} - 1
&& \ell \in L,  k, k' \in K \label{eq_full_alt:force_conflict4} \\
\intertext{\color{blue} Domains of the variables}
& \alpha_e \in \{0,1\} && \hspace*{-2.0cm} \EdgeIdx = \{v, v'\} \in E \\
& a_{\NodeIdx} \in \{0, 1\}  && \hspace*{-2.0cm} v \in \NodeSet \\
& a_{\ell} \in \{0, 1\} && \hspace*{-2.0cm} \LinkIdx \in \LinkSet \\
& {\varphi}_{k \ell} \in \{0,1\} && \hspace*{-2.0cm} \LinkIdx \in \LinkSet, k \in K \\
& \psi_{k \LinkIdx} \in \{0,1\} && \hspace*{-2.0cm} \LinkIdx \in \LinkSet, k \in K \\
& x_k \in \{0, 1\} && \hspace*{-2.0cm} k \in K \\
& \theta_{kk'} \in \{0,1\} 
&& \hspace*{-2.0cm} k, k' \in K\\
& \omega_{k k' \ell} \in \{0,1\} 
&&  \hspace*{-2.0cm} k, k' \in K, \ell \in L.
\label{eq:conflict4}
\end{alignat}

Constraints \eqref{eq:subtour} are the classical subtour elimination in order to guarantee an acyclic graph structure (i.e., a supporting tree for the f-subnet under construction) \cite{nem99}.
Constraints \eqref{eq:single_tree} force the generation of a single f-subnet.
Constraints \eqref{eq:edge_exists_if} and \eqref{eq:node_belongs_if} guarantee the consistency between node and edge variables: an edge is used in the undirected tree if and only if its two endpoints belong to it.
Constraints \eqref{eq:links_vs_edge} take care of consistency between variables: guarantee that if a link is used, its associated edge belongs to the undirected tree. 

The next block of constraints \eqref{eq:request_link1}-\eqref{eq:flow2} take care of the routing of the requests, including the links hosting the unfiltered channels. 
The constraints \eqref{eq:request_link1} and \eqref{eq:request_link2} ensure that if a request is routed over link $\ell$, then that link belongs to the output f-subnet and vice versa, if a link belongs to the output f-subnet structure, then at least one routing uses it.
Constraints \eqref{eq:loop} prevent the use of both $\ell$ and $\overline{\ell}$ (the link in the opposite direction of $\ell$) in the routing of a given request.
Constraints \eqref{eq:flow1}-\eqref{eq:flow2} are flow constraints in order to take care of the "filtered" part of the  routing of the requests.

The next two sets of constraints \eqref{eq:unfiltered1}-\eqref{eq:unfiltered2} are for taking care of the propagation of the unfiltered channels. 
Constraints \eqref{eq:unfiltered1} enforce for every request $k$ and node different from its source node ($s_k$), that if none of its incoming is on the route or broadcast effect of request $k$, then none of the outgoing links can be used for either the routing or the broadcast effect of request $k$.
Constraints \eqref{eq:flowPsi1}-\eqref{eq:flowPsi2} are flow constraints in order to take care of the routing of request $k$ goes through $\LinkIdx$ between its source and destination. Constraints \eqref{eq:flowPsi3} impose no wavelength assignment on the incoming links of the source and on the outgoing links of the destination for the "filtered" routing of $k$.

Constraint \eqref{eq:variables_psy} is the reach constraint, for each request the routing distance between source and destination must not exceed a maximum distance (1,500 km).

Constraints \eqref{eq:phi_psi} define the relation between variables $\varphi_{k \ell}$ and $\psi_{k \ell}$.
Constraints \eqref{eq:conflict1}-\eqref{eq:conflict3} are conflict wavelength constraints either $k$ and $k'$ share a link between their source and destination nodes, see \eqref{eq:conflict1}, or 
one of the requests is routed to the broadcast part of the other request, see \eqref{eq:conflict2} and \eqref{eq:conflict3}.
Constraints \eqref{eq_full_alt:force_conflict1}-\eqref{eq_full_alt:force_conflict4} identify the no conflict case, i.e., when the routes between their source and destination are not overlapping, and then none of the requests is routed to the broadcast part of the other request.


\subsection{PP$^{\textsc{h}}_{\textsc{color}}$: Relaxed Wavelength Pricing Problem}
\label{sec:HW_Pricing}

Wavelength pricing problem aims to find a new wavelength configuration which could potentially improve the objective of the continuous relaxation of the restricted master problem.

The output of this pricing problem contains a set of requests which could share the same wavelength. \\

\noindent
\textbf{Variables} \\
$\beta_k = 1$ if request $k$ uses the wavelength associated with the current wavelength pricing problem, 0 otherwise, for all $k \in K$ \\
$\alpha_{k k'} = \beta_k \> \beta_{k'}$. It is equal to 0 if $k$ and $k'$ cannot be assigned the same wavelength, 0 otherwise. \\

\noindent
Objective:
\begin{equation}
\min \>  1 
- \sum\limits_{k \in K} \dual_k^{(\ref{eq:one_request_at_least_one_color})} \beta_k
+ \sum\limits_{k, k' \in K} \dual_{k k'}^{\eqref{eq:conflict_color}} \alpha_{k k'}
\label{eq:color_obj}
\end{equation}
subject to:
\begin{alignat}{2}
& \beta_k + \beta_{k'} \leq 1 && \nonumber \\
& \qquad \text{if } \sum\limits_{\subnetIdx \in \subnetSet} \theta^{\subnetIdx}_{k k'} z^{\subnetIdx} > 0 \quad 
&& k, k'\in K \label{eq7:conflict}\\
& \alpha_{k k'} \leq \beta_k && k, k'\in K \label{eq7:conflict2}\\
& \alpha_{k k'} \leq \beta_{k'} && k, k'\in K \label{eq7:conflict3}\\
& \beta_k + \beta_{k'} \leq \alpha_{k k'} + 1 && k, k'\in K \label{eq7:conflict4}\\
& \alpha_{k k'} \in \{0,1\} && k, k'\in K \label{eq7:def_alpha}\\
& \beta_k \in \{0,1\} && k \in K. \label{eq7:def_beta}
\end{alignat}
Constraints \eqref{eq7:conflict} identify wavelength conflicts between two requests, and consequently make sure that wavelength conflicting requests are not assigned the same wavelength.
Constraints \eqref{eq7:conflict2}-\eqref{eq7:conflict4} are the linearization of $\alpha_{k k'} = \beta_k \> \beta_{k'}$.

\subsection{PP\textnormal{$^{\textsc{e}}_{\textsc{color}}$}: Exact Coloring Pricing Problem}
\label{sec:EW_Pricing}

The exact wavelength pricing problem is the heuristic one with the omission of constraints \eqref{eq7:conflict}.


\section{Solution Scheme}
\label{sec:sol_scheme}

We now describe the solution process of the Master problem described in Section \ref{sec:Master_Pb}, in coordination with the pricing problems PP$_{\subnetIdx}$, PP$^{\textsc{h}}_{\textsc{color}}$ and PP$^{\textsc{e}}_{\textsc{color}}$, described in Sections \ref{sec:FSN_Pricing}, \ref{sec:HW_Pricing} and \ref{sec:EW_Pricing}.


\subsection{Column Generation and ILP Solution}

Column Generation method is a well-known technique for solving efficiently large-scale Linear programming problems \cite{lub05}. In order to derive an ILP solution, we can then use either a branch-and-price method \cite{bar98} or heuristic methods \cite{sad19}, the accuracy of which can be estimated.

We first discuss the column generation technique and the novelty we introduced with our mathematical formulation. It allows the solution of the linear relaxation of the 
Master Problem, i.e., \eqref{eq:obj_Master}-\eqref{eq:variable_x}. 
 In order to do so, we first define the so-called Restricted Master Problem (RMP), i.e., the Master Problem with a very limited of variables or columns, and a so-called Pricing Problem, i.e., a configuration generator. Here, in our modelling, contrarily to the classical Dantzig-Wolfe decomposition, we consider two different types of pricing problems, the FSN one, and the wavelength one.
 
 The objective function of the Pricing Problems is defined by the Reduced Cost of the decision variable associated with the newly generated configuration: if its minimum value is negative, the addition of the latter configuration will allow a decrease of the optimal LP value of the current RMP, otherwise, if its minimum value is positive for both pricing problems (i.e., the LP optimality condition), the optimal LP value ($z_{\LP}^{\star}$) of problem \eqref{eq:obj_Master}-\eqref{eq:variable_x} has been reached. In other words, the solution process consists in solving the RMP and the pricing problems alternatively, until the LP optimality condition is satisfied.
 
 \begin{figure}[htb]
  \includegraphics[width=.45\textwidth]{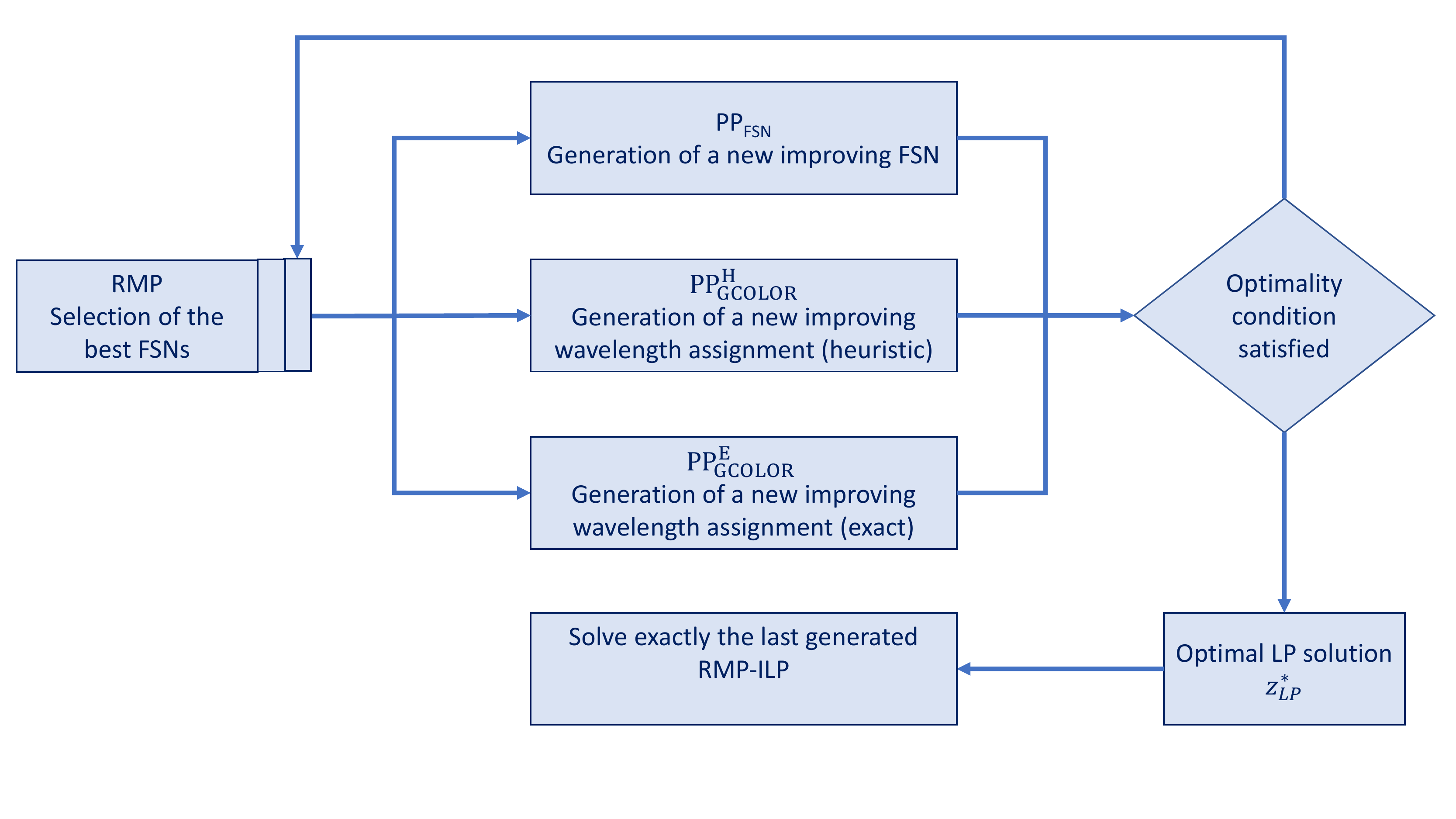} 
  \caption{Flowchart: column generation with three different pricing problems}
  \label{fig:flowchart}
\end{figure} 
 
 Once the optimal LP solution of the MP has been reached, we can then derive an ILP solution, i.e., a selection of pairwise link disjoint filterless sub-networks, which provision all the requests, jointly with a wavelength assignment for all the requests. It is done with the ILP solution of the last RMP in which domains of the variables has been changed from continuous to integer requirements. Denote by ${\tilde z}_{\ILP}$ the value of that ILP solution: it is not necessarily an optimal ILP solution, but is guaranteed to have an accuracy not larger than 
 $$\varepsilon = ({\tilde z}_{\ILP} - z^{\star}_{\LP}) /\ {z_{\LP}^{\star}}. $$


\subsection{Detailed Flowchart and Algorithm}

After outlining the general process of a solution process with a column generation algorithm in the previous section, and on the derivation of an $\varepsilon$-optimal ILP solution, we now provide the detailed flowchart of our solution process in Figure \ref{fig:flowchart}. Solution alternance of the three pricing problems is sought until the LP optimality condition is satisfied, i.e., no configuration with a negative reduced cost can be found. Different strategies can be defined for this alternance, and we describe below the one which gave the best results, among the ones we tested.

\begin{enumerate}
\item \label{step:one} Generate an initial solution with a single FSN supported by a spanning tree, and a first set of wavelength assignment. We use the algorithm of Welsh-Powell \cite{wel67} to minimize the number of wavelengths, using a wavelength conflict graph on which we minimize the number of colors.
\item \label{step:two} Apply the column generation algorithm in order to solve the linear relaxation of model {\ModelName}: 
   \begin{enumerate}
        \item Solve restricted master problem with current FSN and wavelength assignment configurations \label{step:master}  
        \item Solve pricing problem PP$_{\subnetIdx{}}$. If it generates an improving FSN configuration, add it to the RMP and return to Step \ref{step:master} \label{step:pricing_FSN}
        \item Solve pricing problem PP$^{\textsc{h}}_{\textsc{color}}$. If it generates an improving wavelength configuration, add it to the RMP and return to Step \ref{step:master} \label{step:h_pricing_color}
        \item Solve pricing problem PP$^{\textsc{e}}_{\textsc{color}}$. If it generates an improving wavelength configuration, add it to the RMP and return to Step \ref{step:master} \label{step:e_pricing_color}
   \end{enumerate} 
\item As the continuous relaxation of the Master Problem has been solved optimally, solve the last generated restricted master problem with integer requirements for the variables, derive an ILP solution.
\end{enumerate}


\section{Numerical Results}
\label{sec:results}

We report now the performance of our proposed mathematical model and algorithm, and compare the quality of its solutions with those of Tremblay \textit{et al.} \cite{tre13}.


\subsection{Data Sets}

We use the Italy, California and Germany topologies as in Tremblay \textit{et al.} \cite{tre13} (see \cite{arc08} for distances), as well as the Cost239 and USA topologies, using the link distances of \cite{had17} and \cite{sndlib}, respectively. We recall their characteristics in Table \ref{tab:dataSets}. We consider unit uniform demands, as in \cite{tre13}, unless otherwise indicated.
\begin{table}[htb]
\caption{Data set characteristics}
\label{tab:dataSets}
\centering
    \begin{tabular}{|p{2cm}|p{1.5cm}| p{1.5cm}|}
    \hline
    Networks & \# nodes & \# nodes  \\
    \hline  
    Italy      & \hspace*{.3cm} 10 & \hspace*{.3cm} 15  \\
    California & \hspace*{.3cm} 17 & \hspace*{.3cm} 20 \\    
    Germany17  & \hspace*{.3cm} 17 & \hspace*{.3cm} 26 \\
    Cost239    & \hspace*{.3cm} 11 & \hspace*{.3cm} 26 \\
    USA        & \hspace*{.3cm} 12 & \hspace*{.3cm} 15  \\
    \hline         
    \end{tabular}
\end{table}
\begin{table*}[hbt]
    \caption{Network parameters for filterless solutions}
    \label{tab:comparison}
    \centering
    \begin{small}
    \begin{tabular}{|l|cr|cccr|cccr|cccc|}
    \cline{2-15}
    \multicolumn{1}{c|}{} 
    & \multicolumn{2}{c|}{\multirow{2}{*}{Tremblay \textit{et al.} \cite{tre13}}} 
    & \multicolumn{12}{c|}{Model \ModelName} \\
    \cline{4-15} 
    \multicolumn{1}{c|}{} 
    & \multicolumn{2}{c|}{}
    & \multicolumn{4}{c|}{a single tree}
    & \multicolumn{4}{c|}{two trees}
    & \multicolumn{4}{c|}{three trees}\\
    \cline{2-15} 
    \multicolumn{1}{c|}{} 
    &\multirow{2}{*} {\# Trees} & \multirow{2}{*}{W}     
    & \multirow{2}{*}{$z^{\star}_{\textsc{lp}}$} & \multirow{2}{*}{W} & \multicolumn{2}{c|}{Max Load} 
    & \multirow{2}{*}{$z^{\star}_{\textsc{lp}}$} & \multirow{2}{*}{W} & \multicolumn{2}{c|}{Max Load}
    & \multirow{2}{*}{$z^{\star}_{\textsc{lp}}$} & \multirow{2}{*}{W} & \multicolumn{2}{c|}{Max Load}\\
    \multicolumn{1}{c|}{} 
    &  & & & & R & N 
    &  & & R & N 
    &  & & R & N \\
    \hline
    Italy & 2 & \phantom{1}25 
          & \phantom{1}41.0 & \phantom{1}41 & \phantom{1}9 & 32
          & \phantom{1}20.0 & \phantom{1}23 & \phantom{1}7 &16 &- &- &- &- \\
    California & 3 & 120      
               & 125.6 & 126 &16 &  110
          & \phantom{1}117.4 & 122 & 16 &106  & \phantom{1}113.4& 120&16 & 104\\
    Germany17 & 2 & \phantom{1}88  
              & 120.5 & 125 &16 & 109 
              &\phantom{1}62.3 & \phantom{1}73 & \phantom{1}6 &  67&- &- &- &- \\
    Cost239 & - & - 
            & \phantom{1}49.3 & \phantom{1}51 & 10 & 41 
            & \phantom{1}22.7 & \phantom{1}28 & \phantom{1}6 & 22 &\phantom{1}15.7 & \phantom{1}25 & \phantom{1}7 &18 \\
    USA & - & -  
        & \phantom{1}61.0 & \phantom{1}61 & 11& 50 
        & \phantom{1}41.2 & \phantom{1}53 & \phantom{1}7 & 46 &- &- &- &- \\ 
    \hline
    \end{tabular}
    \end{small}
\end{table*}


\subsection{Performance of the Proposed \textnormal{\ModelName} Model}

We tested our model/algorithm on the same data sets as Tremblay \textit{et al.} \cite{tre13}. 
Results are summarized in Table \ref{tab:comparison}. 

Available numerical results of \cite{tre13} are reported in columns 2 and 3, while results or our {\ModelName} model are reported in the remaining columns.
We provide the results of the {\ModelName} model in Columns 4 to 7 (2 trees for Italian and Germany17, 3 trees for California), for the first three data sets. 

\begin{figure*}
  \centering
  \subfigure[Italy]{\includegraphics[width=.3\textwidth]{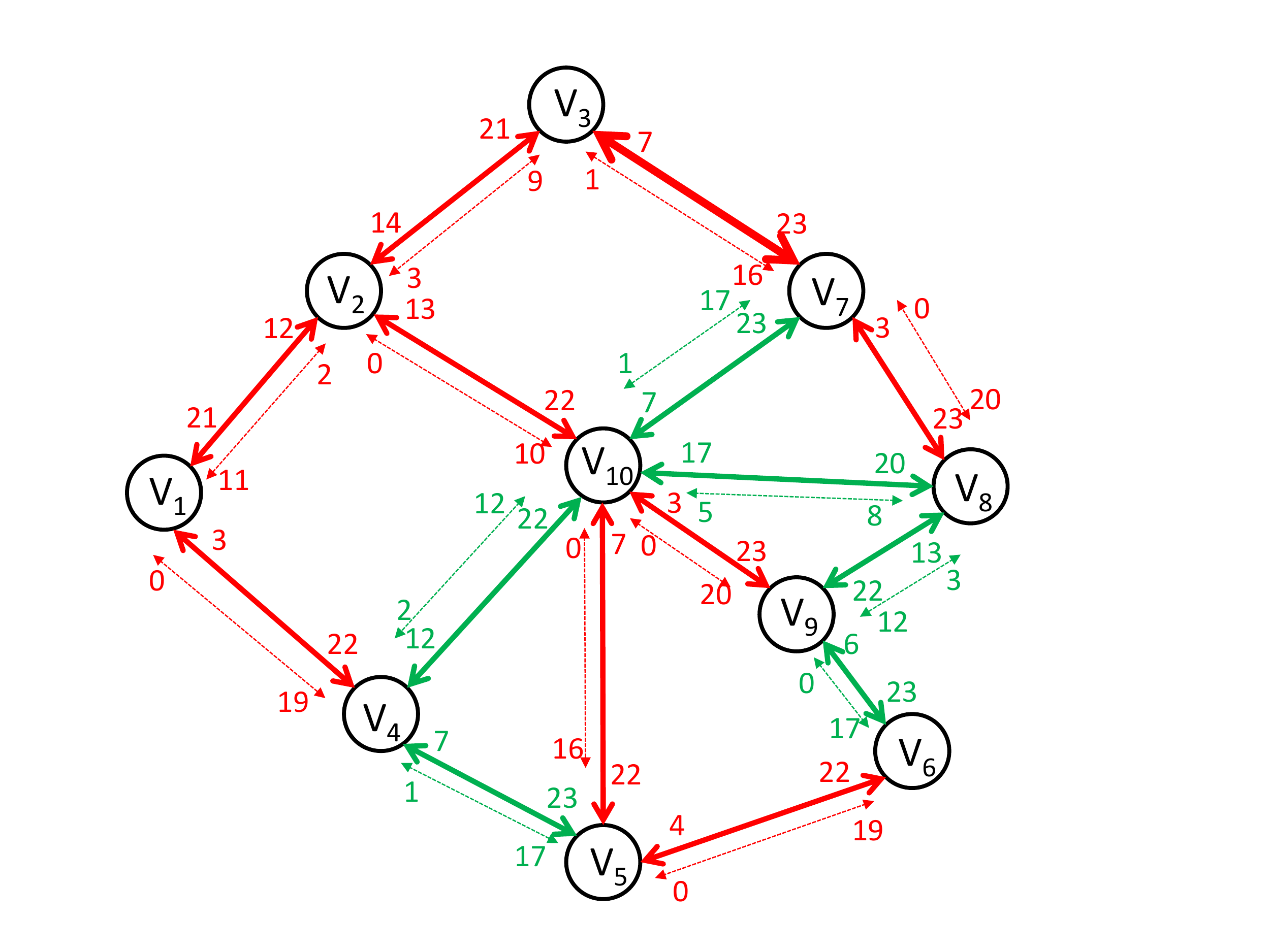} }
  \subfigure[California]{\includegraphics[width=.3\textwidth]{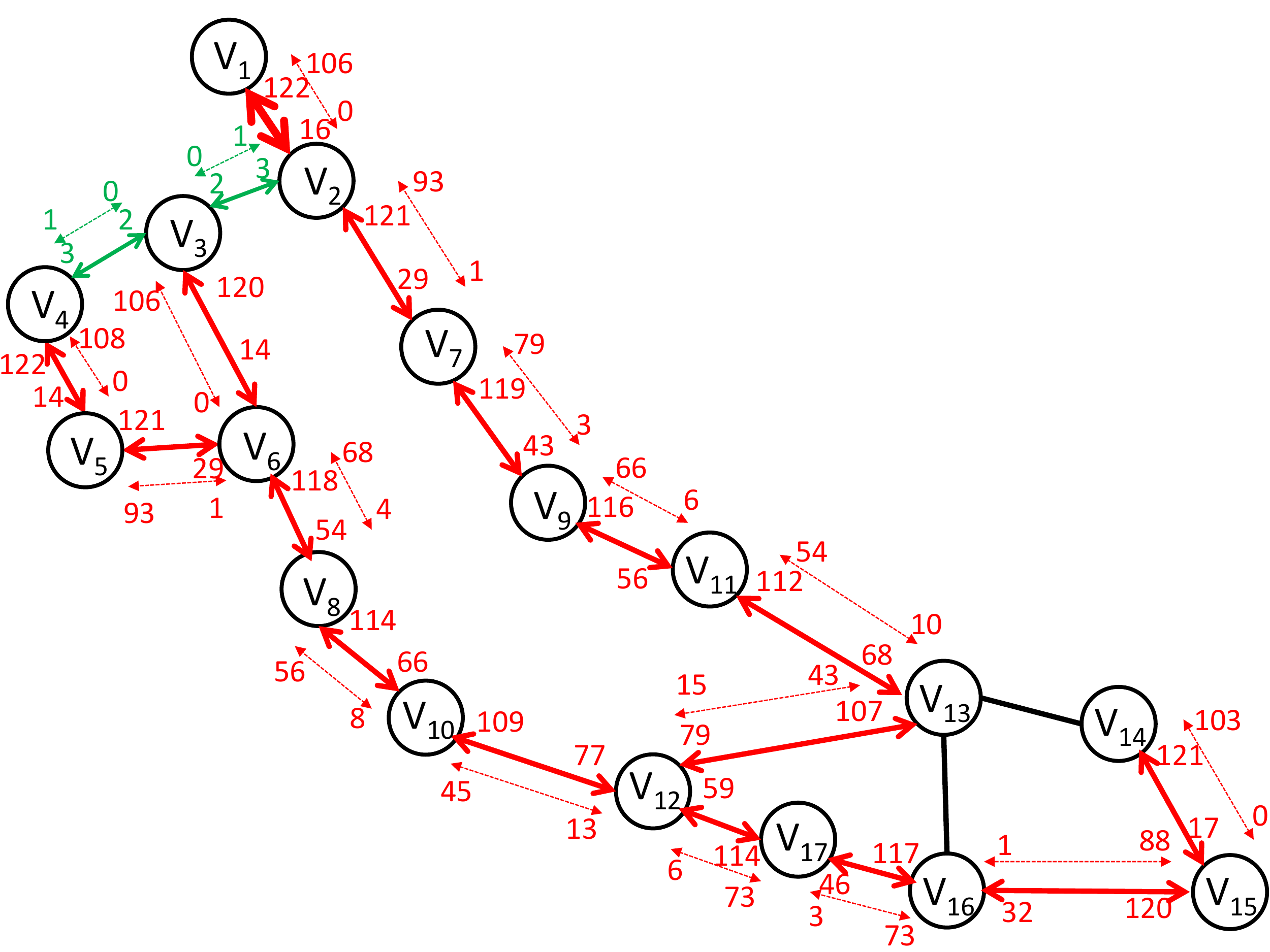}}
  \subfigure[Germany17]{\includegraphics[width=.3\textwidth]{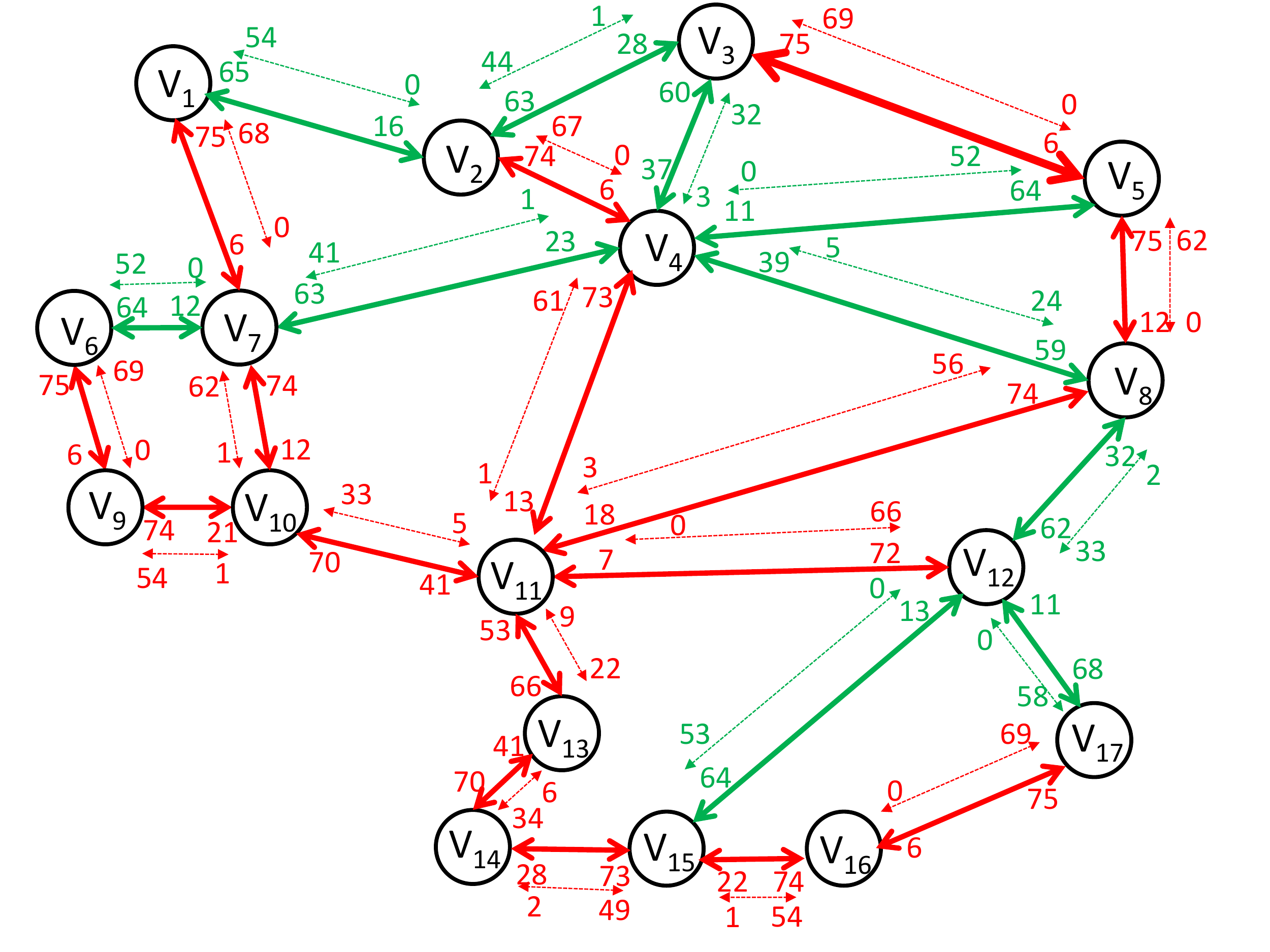}}
  \caption{Two filterless sub-networks solutions with optimized trees compared to those of \cite{tre13}}
  \label{fig:sol3}
\end{figure*}

\begin{figure*}
    \centering
    \subfigure[Uniform Traffic (272 unit requests, 1 request per node pair)]
        {\includegraphics[width=0.45\linewidth]{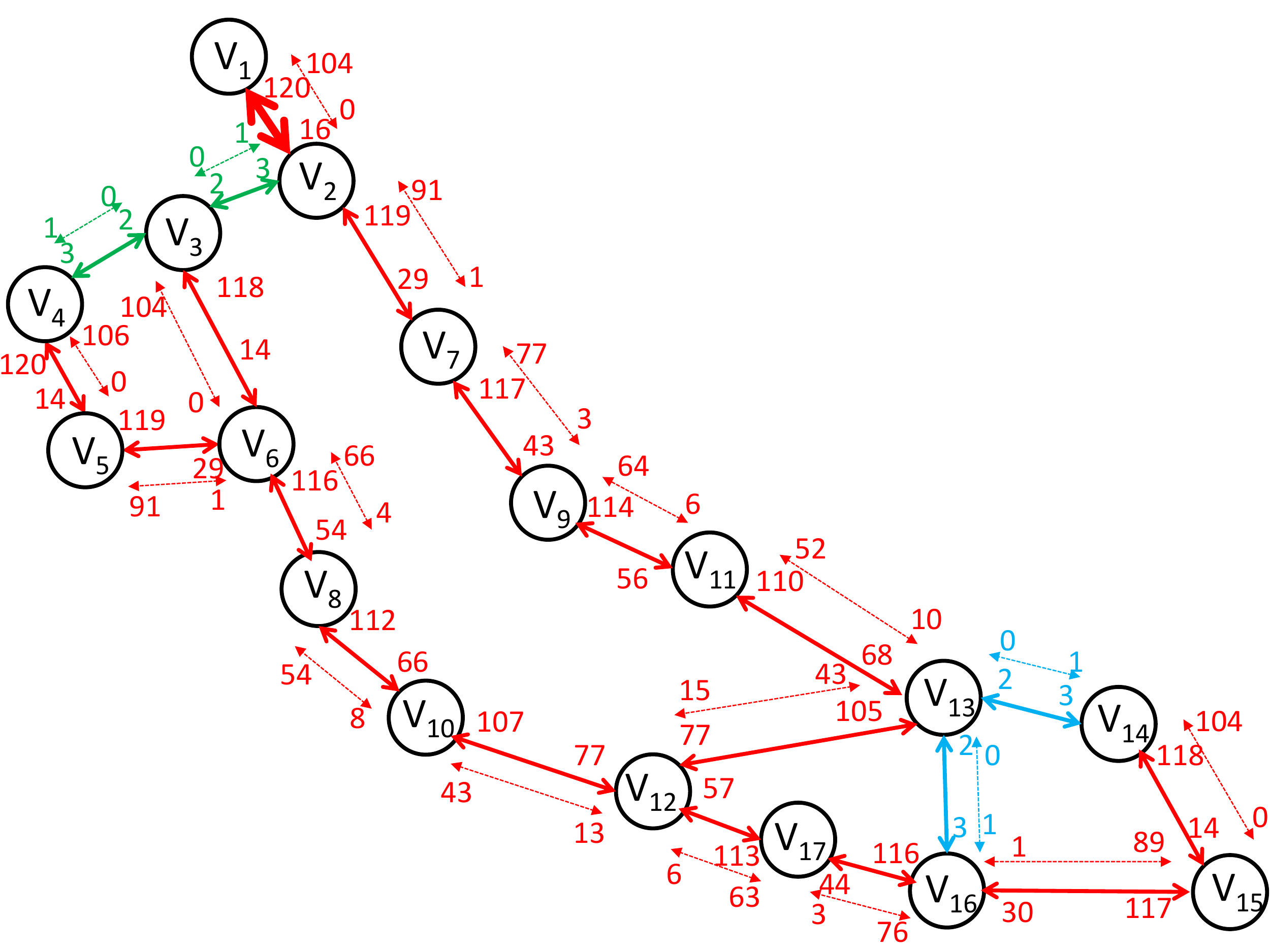}\label{fig:5a}}
    \subfigure[Non Uniform Traffic (338 unit requests, several requests per node pair)]
        {\includegraphics[width=0.45\linewidth]{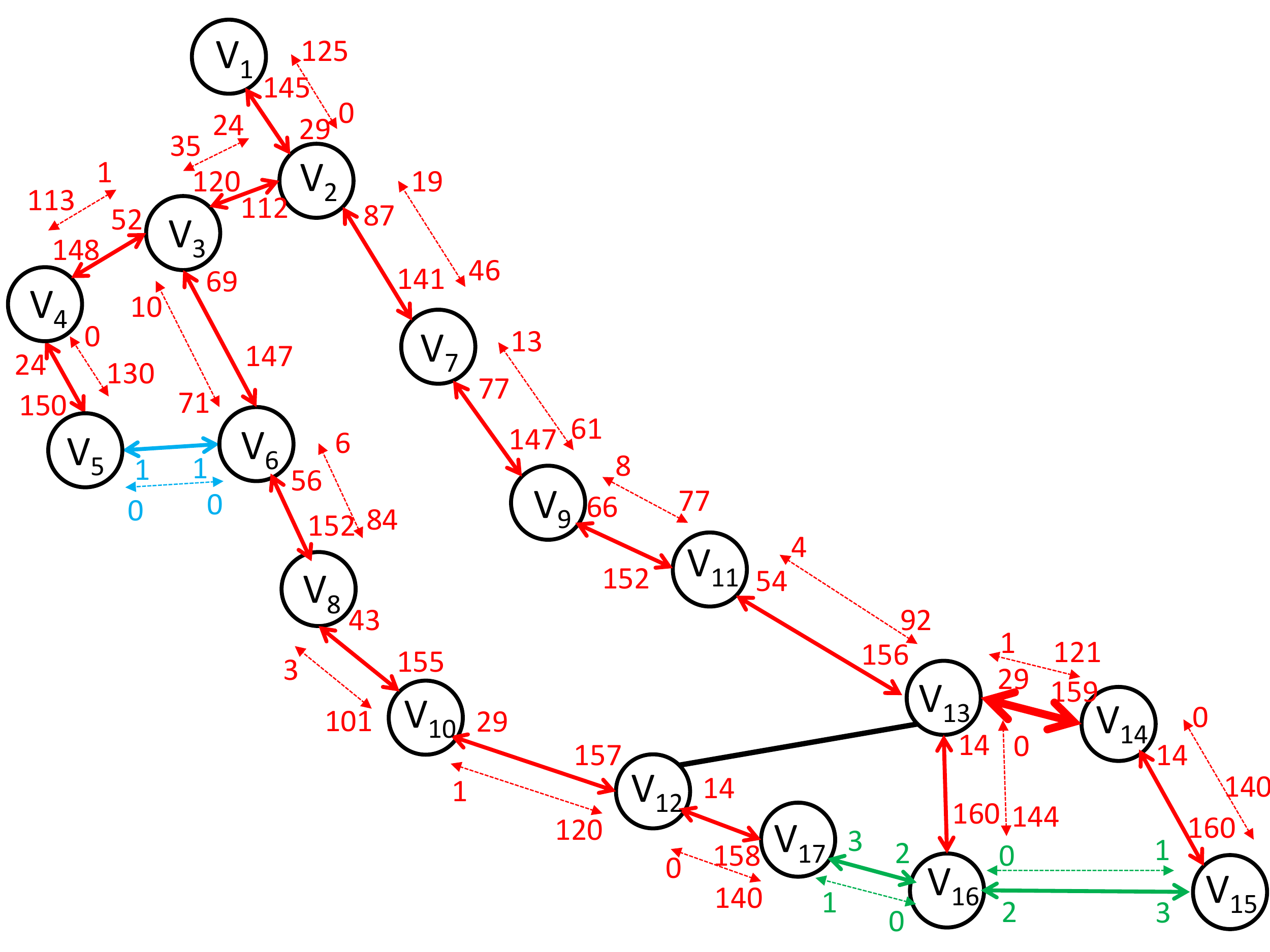}       \label{fig:5b}}
    \caption{Three filterless sub-networks on the California network}
    \label{fig:fig5}
\end{figure*}

We observe that we improve the values of Tremblay \textit{et al.} \cite{tre13} when using the same 2 supporting trees for 2 for the FSNs of two data sets,
i.e., 23 wavelengths instead of 25 for the Italy network,
and 73 wavelengths instead of 88 for Germany network
In Figure \ref{fig:sol3}, we provide the details of the solutions of Model {\ModelName} with different trees than those of 
 \cite{tre13}. Numbers on the links indicate the number of wavelengths, by differentiating between  "filtered" and "unfiltered" wavelengths.

As we did not develop a branch-and-price algorithm and as we limited the number of iterations in the column generation algorithm, the solutions of Model {\ModelName} are not necessarily optimal, except when $z^{\star}_{\textsc{lp}}$ is equal to the ILP optimal value. We then observe that several solutions with one tree are optimal, i.e., those for the Italy, California and USA networks. 

We also provided results with one tree (optimized selection): we observe that the number of wavelengths for one tree is increasing over the required number for 2 or 3 trees. 

Lastly, in Figure \ref{fig:fig5}, we provide results with three f-subnets for the California network, first with uniform traffic in Figure \ref{fig:5a}, and then with non uniform traffic in Figure \ref{fig:5b} (number of requests randomly generated between 1 and 3 for each node pair). It is interesting to observe that the optimal solution for the non uniform traffic does not use one link, and that while one f-subnet is always a spanning, the two other f-subnets are very small, most likely due to the characteristics of the California topology. The bold red  arrow indicates the link with the largest number of wavelengths.


\section{Conclusions}

This paper presents a one step decomposition model and algorithm, which can solve exactly the filterless network design problem. 
While computational times are sometimes quite high (i.e., sometimes a couple of hours), we hope to be soon able to improve further the modelling and algorithm in order to provide more scalable solutions. However, this is already a very significant first step towards the exact design of filterless optical networks.

\section*{Acknowledgment}

B. Jaumard has been supported by a Concordia University Research Chair (Tier I) on the Optimization of Communication Networks and by an NSERC (Natural Sciences and Engineering Research Council of Canada) grant.

\bibliographystyle{IEEEtran}


\end{document}